\documentclass[prd,aps,showpacs,nofootinbib]{revtex4}%
\usepackage{graphicx}
\usepackage{amsmath}
\usepackage{amsfonts}
\usepackage{amssymb}%
\newcommand{\be}{\begin{equation}}
\newcommand{\ee}{\end{equation}}
\newcommand{\bq}{\begin{eqnarray}}
\newcommand{\eq}{\end{eqnarray}}
\begin{document}
\title{Non-relativistic electron-electron interaction in a Maxwell-Chern-Simons-Proca
model endowed with a timelike Lorentz-violating background}
\author{Manoel M. Ferreira Jr and M\'{a}rcio S. Tavares}
\affiliation{Universidade Federal do Maranh\~{a}o (UFMA), Departamento de F\'{\i}sica,
Campus Universit\'{a}rio do Bacanga, S\~{a}o Luiz - MA, 65085-580 - Brasil}

\begin{abstract}
A planar Maxwell-Chern-Simons-Proca model endowed with a Lorentz-violating
background is taken as framework to investigate the electron-electron
interaction. The Dirac sector is introduced exhibiting a Yukawa and a minimal
coupling with the scalar and the gauge fields, respectively. The the
electron-electron interaction is then exactly evaluated as the Fourier
transform of the M\"{o}ller scattering amplitude (carried out in the
non-relativistic limit) for the case of a purely time-like background. The
interaction potential exhibits a totally screened behavior far from the origin
as consequence of massive character of the physical mediators. The total
interaction (scalar plus gauge potential) can always be attractive, revealing
that this model may lead to the formation of electron-electron bound states.

\end{abstract}
\email{manojr@ufma.br, plasmatavares@yahoo.com.br}

\maketitle

\section{Introduction}

In 1990, Carroll-Field-Jackiw \cite{Jackiw} have proposed a version of the
Maxwell electrodynamics corrected by a Chern-Simons-like term $\left(
\epsilon_{\mu\nu\alpha\beta}V^{\mu}A^{\nu}F^{\alpha\beta}\right)  $ in order
to incorporate a Lorentz-violating background $\left(  V^{\mu}\right)  $ into
the usual electrodynamics. This term implies a modified theory in which
photons with different polarizations propagate with distinct velocities
(birefringence). Some years later, Colladay \& Kostelecky \cite{Colladay1}%
-\cite{Colladay2} have constructed an extension of the Minimal Standard Model,
the Extended Standard Model (SME), in which Lorentz-violating tensor terms,
stemming from a spontaneous symmetry breaking (SSB) of a more fundamental
theory (defined at the Planck scale) are properly incorporated in all
interaction sectors. The construction of the SME was in part motivated by
works demonstrating the possibility of Lorentz and CPT spontaneous violation
in the context of string theory \cite{Kostelecky1}-\cite{Kostelecky3}.
Recently, the SME has motivated innumerous interesting works \cite{Chung}%
-\cite{PRD2}. One of the most remarkable controversies involving Lorentz
violation deals with the radiative generation of the Carroll-Field-Jackiw term
from the integration on the fermion fields \cite{Chung}-\cite{Brito}. Lorentz
violating theories investigations have also been concerned with the
consistency aspects of the Carroll-Field-Jackiw electrodynamics \cite{Adam}%
-\cite{Lehnert}, study of synchrotron radiation, electrostatics and
magnetostatics in Lorentz-violating electrodynamics \cite{Urrutia}%
-\cite{Bailey}, influence of Lorentz violation on the Dirac equation
\cite{Lehnert2}-\cite{Fernando1}, CPT-probing experiments \cite{Bluhm1}%
-\cite{Fernando2}, Cerenkov radiation \cite{Lehnert3}, and general aspects
\cite{Andrianov1}-\cite{Gaete}.

A theoretical model which provides an attractive electron-electron interaction
could work, in principle, as a good framework to properly address the
electron-electron pairing in planar systems. In fact, if an attractive
electron-electron interaction is obtained in the context of a particular
model, it may be seen as a first connection between such theoretical models
and the attainment of electron pairing. In practice, this interplay has begun
with the application of the Maxwell-Chern-Simons (MCS) theory \cite{MCS}%
-\cite{Georgelin} for evaluating the electron-electron interaction in a planar
model. However, it was soon established that the MCS model does not imply an
attractive interaction for small topological mass $\left(  s<<m_{e}\right)  ,$
regime compatible with low-energy excitations. Currently it is well known that
by including the Higgs sector \cite{Tese}, \cite{Tese2}, an attractive
interaction can be got, assured it is suitably coupled to the fermion field by
a quartic order term - $\overline{\psi}\psi\phi^{\ast}\phi$ (that gives rise
to the Yukawa coupling with the Higgs field after SSB). On the other hand, it
has been recently verified that the MCS theory may also yield an attractive
$e^{-}e^{-}$ potential (in the absence of Higgs sector) provided it is
considered in the presence of a fixed Lorentz-violating background.
Specifically, it has been evaluated the electron-electron interaction
potential in the context of a planar Lorentz-violating Maxwell-Chern-Simons
model (arising from the dimensional reduction of the
Maxwell-Carroll-Field-Jackiw model). Such calculation was carried out both for
the case of a purely timelike background \cite{PRD1} and for a purely
spacelike background \cite{PRD2}, leading to interacting potentials with a
well (attractive) region.

Some years ago, it was argued that there is a relation between Lorentz
violation and noncommutativity \cite{Carroll}. The introduction of
noncommutativity in the MCS model \cite{Ghosh} has appeared as a new mechanism
able to provide $e^{-}e^{-}$ attraction. In fact, the noncommutative extension
of the minimal MCS model has shown to be a suitable framework to provide an
attractive electron-electron potential. Specifically, this model yields the
same interaction potential attained by Georgelin \& Wallet \cite{Georgelin}
considering a non-minimal Pauli magnetic coupling. This puts in evidence the
relevance of non usual mechanisms for the attainment of electron-electron
attractiveness. It should be also mentioned that noncommutative Chern-Simons
theories have been applied successfully to describe properties of planar Hall
systems \cite{NC1}-\cite{NC3}, one of the points of clear connection of
noncommutativity with condensed matter physics. The large number of
applications of noncommutativity to condensed matter physics and the general
relation between these mechanisms indicate that applications of Lorentz
violation to condensed matter systems should be a sensible and feasible issue
as well.

Lorentz violation in the presence of the Higgs sector and spontaneous symmetry
breaking (SSB)\ was first investigated in the context of the 4-dimensional
Abelian-Higgs Carroll-Field-Jackiw electrodynamics \cite{Belich1}. This model,
by means of a dimensional reduction procedure, has originated a planar
electrodynamics composed of the MCS sector with a Higgs field, coupled to a
massless Klein-Gordon mode - $\varphi$ (stemming from the dimensional
reduction, $\varphi=A^{(3)}$) and to the Lorentz-violating background
($v^{\mu})$ \cite{Manojr3}. The consistency of this model was also set up (it
turned out to be totally causal, unitary and stable). Once SSB takes place,
the gauge and the Klein-Gordon fields acquire mass, giving rise to a MCS-Proca
electrodynamics coupled to the Lorentz-violating background. Such a model was
already used to perform an investigation concerned with condensed matter
physics: the study of vortex-like solutions in a planar Lorentz-violating
environment \cite{Vortex}. As a result, it was shown that it provides charged
vortex solutions that recover the usual Nielsen-Olesen configuration in the
asymptotic regime.

In the present work, the aim is to investigate the electron-electron
interaction in the context of the Abelian-Higgs Lorentz-violating planar model
previously defined, another issue with possible connection with condensed
matter physics. The Lagrangian of ref. \cite{Manojr3} does not stand for the
most general neither the most simple model to perform such a task. However,
there are two good reasons to adopt it: its consistency has been already
established\footnote{This choice assures that the adopted model, here analyzed
at tree-level, may be consistently quantized in other specific applications.};
it is expected that it will provide shielded versions of the MCS
Lorentz-violating potentials derived in ref. \cite{PRD1} (which present a
logarithmically asymptotic behavior). Having stated the gauge model, the Dirac
sector is properly incorporated in it by exhibiting the minimal and the Yukawa
couplings with the gauge ($A_{\mu})$ and the scalar ($\varphi)$ fields,
respectively. One then proceeds to carry out the electron-electron potential
for the case of a purely timelike background and to discuss its possible
attractiveness. The procedure is much similar to the one adopted in refs.
\cite{PRD1}, \cite{PRD2}: starting from a known planar Lagrangian, the
M\"{o}ller scattering amplitudes (for the gauge and scalar intermediations)
are constructed; next, its Fourier transforms are evaluated, leading to the
interacting potentials. In the present case, the result is a totally screened
potential \textbf{(}due to the presence of the Higgs sector\textbf{)} composed
of the sum of a scalar and a gauge contribution.\ This potential exhibits an
attractive behavior even in the presence of the centrifugal barrier and the
$A^{2}-$gauge invariant term, thereby confirming its attractive character
(even under a more rigorous analysis) and its possible relevance to the
formation of electron-electron pairs in planar systems. The results obtained
here are compared with the one of ref. \cite{PRD1} in order to emphasize the
role played by the Higgs sector: it transforms logarithmically divergent
solutions in entirely shielded ones. This comparison will be accomplished
throughout this work. Another point to be remarked is that the gauge
potentials here derived may be attractive while the ones of refs. \cite{Tese},
\cite{Tese2} are always repulsive, which constitutes a sensitive difference
between the results of these works.

This work is outlined as follows. In Sec. II, the reduced model derived in
ref. \cite{Manojr3}, supplemented by the fermion field, is briefly described.
In Sec. III are presented the spinors which fulfill the two-dimensional Dirac
equation, which, in its turn, are used to evaluate the M\"{o}ller scattering
amplitude associated with the scalar and gauge intermediations. In Sec. IV,
the interaction potentials are evaluated by performing the Fourier transform
of the scalar and gauge scattering amplitudes. The results are properly
discussed. In Sec.V are presented the concluding remarks.

\section{The Planar Lorentz-violating model}

At first, one ought to present the planar model which sets up the theoretical
framework for the calculations realized in this work. The starting point is
the (1+3)-dimensional Abelian-Higgs Maxwell-Carroll-Field-Jackiw (MCFJ)\ model
\cite{Belich1}, consisting of the MCFJ electrodynamics supplemented with the
Higgs sector.%

\begin{equation}
\mathcal{L}_{1+3}=\{-\frac{1}{4}F_{\hat{\mu}\hat{\nu}}F^{\hat{\mu}\hat{\nu}%
}+\frac{1}{4}\varepsilon^{\hat{\mu}\hat{\nu}\hat{\kappa}\hat{\lambda}}%
v_{\hat{\mu}}A_{\hat{\nu}}F_{\hat{\kappa}\hat{\lambda}}+(D^{\widehat{\mu}}%
\phi)^{\ast}D_{\widehat{\mu}}\phi-V(\phi^{\ast}\phi)+A_{\hat{\nu}}J^{\hat{\nu
}}\}, \label{action1}%
\end{equation}
where $v_{\hat{\mu}}$ is the Lorentz-violating fixed background, $\hat{\mu}$
runs from $0$ to $3,$ $D_{\widehat{\mu}}\phi=(\partial_{\widehat{\mu}%
}+ieA_{\widehat{\mu}})\phi$ is the covariant derivative and $V(\phi^{\ast}%
\phi)=m^{2}\phi^{\ast}\phi+\lambda(\phi^{\ast}\phi)^{2}$ represents the scalar
potential responsible for SSB ($m^{2}<0$ and $\lambda>0).$ In a previous work,
this model had undergone a dimensional reduction procedure in which the third
spatial coordinate is frozen, implying: $\partial_{_{3}}\chi\longrightarrow0.$
At the same time, the third component of the vector potential becomes a scalar
field, $A^{(3)}\rightarrow\varphi,$ whereas the third component of background
becomes the topological mass:$\ v^{(3)}\longrightarrow$ $s.$ This process
yields a Lorentz-violating planar model incorporating the Higgs sector
\cite{Manojr3}, given as below:%

\begin{align}
\mathcal{L}_{1+2}  &  =-\frac{1}{4}F_{\mu\nu}F^{\mu\nu}+\frac{1}{2}%
\partial_{\mu}\varphi\partial^{\mu}\varphi+\frac{s}{2}\epsilon_{\mu\nu
k}A^{\mu}\partial^{\nu}A^{k}-\varphi\epsilon_{\mu\nu k}v^{\mu}\partial^{\nu
}A^{k}+(D_{\mu}\phi)^{\ast}(D_{\mu}\phi)\nonumber\\
&  -e^{2}\varphi^{2}(\phi^{\ast}\phi)-V\left(  \phi^{\ast}\phi\right)
-A_{\mu}J^{\mu}-\varphi J, \label{Lagrange2}%
\end{align}
where the greek letters now run from $0$ to $2.$ Once the planar
Lorentz-violating model has been established, it is considered the spontaneous
symmetry breaking process which provides mass to the gauge and scalar fields
\cite{Manojr3}. Relying on a tree-level analysis, one should retain only the
bilinear terms, so that the planar Lagrangian takes the form:
\begin{equation}
\mathcal{L}_{1+2}^{broken}=-\frac{1}{4}F_{\mu\nu}F^{\mu\nu}+\frac{1}%
{2}\partial_{\mu}\varphi\partial^{\mu}\varphi-\frac{1}{2}M_{A}^{2}\varphi
^{2}+\frac{s}{2}\epsilon_{\mu\nu k}A^{\mu}\partial^{\nu}A^{k}-\varphi
\epsilon_{\mu\nu k}v^{\mu}\partial^{\nu}A^{k}+\frac{1}{2}M_{A}^{2}A_{\mu
}A^{\mu}-A_{\mu}J^{\mu}-\varphi J, \label{Lagrange3}%
\end{equation}
where $M_{A}^{2}=2e^{2}\langle\phi\phi\rangle,\ $and $\langle\phi\phi
\rangle\ $is the vacuum expectation value of the $\phi-$ field. Here, one
explains the reason for which the Higgs field does not appear in the above
equation: it only keeps high order couplings with the other fields. As it is
well-known, these terms are not taken into account in a tree-level evaluation.
The classical solutions of this planar model were achieved in ref.
\cite{Manojr4}, where the effects of the fixed background on the MCS-Proca
electrodynamics were exhaustively analyzed. It was also reported that the
scalar potential $\left(  A_{0}\right)  $ for a purely timelike background
exhibits an attractive behavior, which may be seen as a cue indicating that a
similar result may be shared by the gauge electron-electron potential to be
evaluated in a dynamic configuration.

Now, it is necessary to introduce the spinor field suitably coupled to the
gauge $\left(  A_{\mu}\right)  $ and scalar $\left(  \varphi\right)  $ ones.
In the absence of sources, the interaction Lagrangian is read as:%

\begin{align}
\mathcal{L}_{1+2}^{broken}  &  =-\frac{1}{4}F_{\mu\nu}F^{\mu\nu}+\frac{1}%
{2}\partial_{\mu}\varphi\partial^{\mu}\varphi-\frac{1}{2}M_{A}^{2}\varphi
^{2}+\frac{s}{2}\epsilon_{\mu\nu k}A^{\mu}\partial^{\nu}A^{k}-\varphi
\epsilon_{\mu\nu k}v^{\mu}\partial^{\nu}A^{k}+\nonumber\\
&  +\frac{1}{2}M_{A}^{2}A_{\mu}A^{\mu}+{\overline{\psi}(i}%
{\rlap{\hbox{$\mskip4.5 mu /$}}D}-m_{e}){\psi}-y\varphi(\overline{\psi}\psi).
\label{L1}%
\end{align}
The term ${\rlap{\hbox{$\mskip4.5
mu /$}}D}\psi\equiv(\rlap{\hbox{$\mskip1
mu /$}}\partial+ie_{3}\rlap{\hbox{$\mskip3 mu /$}}A)\psi$ sets up the minimal
coupling, whereas $\varphi(\overline{\psi}\psi)$ reflects the Yukawa coupling
with the scalar field. The fermion field $\left(  \psi\right)  $\ is a
two-component spinor with up spin-polarization, representing the positive
energy solution of the Dirac equation, $\left(  \gamma^{\mu}p_{\mu}-m\right)
u(p)=0,$\ here written in momentum space. The mass dimension of the fields and
parameters involved in eq. (\ref{L1}) are the following: $\left[
\varphi\right]  =\left[  A^{\mu}\right]  =1/2,\left[  \psi\right]  =1,\left[
s\right]  =\left[  v^{\mu}\right]  =1,\left[  e_{3}\right]  =\left[  y\right]
=1/2;$ it is noticeable that both coupling constants, $e_{3}$ and $y,$ exhibit
$\left[  \text{mass}\right]  ^{1/2}$ dimension, a usual result in (1+2)
dimensions. In ref. \cite{Manojr3}, the propagators of the scalar $\left(
\varphi\right)  $ and gauge $\left(  A_{\mu}\right)  $ fields were properly
evaluated in the following form:
\begin{align}
\langle A^{\mu}\left(  k\right)  A^{\nu}\left(  k\right)  \rangle\text{ }  &
=i\biggl\{-\frac{(k^{2}-M_{A}^{2})}{\boxplus(k)}\theta^{\mu\nu}+\frac
{(k^{2}-M_{A}^{2})\boxtimes\boxplus-\lambda^{2}s^{2}M_{A}^{2}k^{2}}{M_{A}%
^{2}(k^{2}-M_{A}^{2})\boxtimes(k)\boxplus(k)}\omega^{\mu\nu}-\frac{s}%
{\boxplus}S^{\mu\nu}\nonumber\\
&  +\frac{s^{2}k^{4}}{(k^{2}-M_{A}^{2})\boxtimes(k)\boxplus(k)}\Lambda^{\mu
\nu}-\frac{(k^{2}-M_{A}^{2})}{\boxtimes(k)\boxplus(k)}T^{\mu}T^{\nu}%
+\frac{sk^{2}}{\boxtimes(k)\boxplus(k)}\left[  Q^{\mu\nu}-Q^{\nu\mu}\right]
\nonumber\\
&  +\frac{i(v\cdot k)s^{2}k^{2}}{(k^{2}-M_{A}^{2})\boxtimes(k)\boxplus
(k)}\left[  \Sigma^{\mu\nu}+\Sigma^{\nu\mu}\right]  -\frac{is(v\cdot
k)}{\boxtimes(k)\boxplus(k)}\left[  \Phi^{\mu\nu}-\Phi^{\nu\mu}\right]
\biggr\}, \label{Prop_gauge}%
\end{align}%
\begin{equation}
\text{ }\langle\varphi\varphi\rangle\text{ }=i\frac{\boxplus(k)}%
{\boxtimes(k)(k^{2}-M_{A}^{2})}, \label{Prop_scalar}%
\end{equation}
with: $\boxtimes(k)=k^{4}-\left(  2M_{A}^{2}+s^{2}-v\cdot v\right)
k^{2}+M_{A}^{4}-\left(  v\cdot k\right)  ^{2},$ $\boxplus(k)=(k^{2}-M_{A}%
^{2})^{2}-s^{2}k^{2}.$

The projector operators are defined as: $\theta_{\mu\nu}=\eta_{\mu\nu}%
-\omega_{\mu\nu},$\ $\omega_{\mu\nu}=\partial_{\mu}\partial_{\nu}/\square,$
$S_{\mu\nu}=\varepsilon_{\mu\kappa\nu}\partial^{\kappa},$ $Q_{\mu\nu}=v_{\mu
}T_{\nu},T_{\nu}=S_{\mu\nu}v^{\mu},$ $\Lambda_{\mu\nu}=v_{\mu}v_{\nu},$
\ $\Sigma_{\mu\nu}=v_{\mu}\partial_{\nu},$ $\Phi_{\mu\nu}=T_{\mu}\partial
_{\nu},$ while it is adopted the (1+2) metric convention: $\eta_{\mu\nu
}=(+,-,-).$ Naturally, these expressions are essential for the evaluation of
the amplitudes associated with the M\"{o}ller scattering, task developed in
the next section.

\section{The M\"{o}ller Scattering amplitude}

In the context of a low-energy interaction, the Born approximation holds as a
good approximation. Consequently, the interaction potential arises as the
Fourier transform of the two-particle scattering amplitude. Another point is
that, in the case of the nonrelativistic M\"{o}ller scattering, it should be
considered only the direct scattering process \cite{Sakurai} (even for
indistinguishable electrons), since in this limit they recover the classical
notion of trajectory. From eq. (\ref{L1}), one extracts the Feynman rules for
the interaction vertices involving fermions: $V_{\psi\varphi\psi}=iy;V_{\psi
A\psi}=ie_{3}\gamma^{\mu}$. Therefore, the $e^{-}e^{-}$ scattering amplitudes
are read as:
\begin{align}
-i\mathcal{M}_{\varphi}  &  =\overline{u}(p_{1}^{\prime})(iy)u(p_{1})\left[
\langle\varphi\varphi\rangle\right]  \overline{u}(p_{2}^{\prime}%
)(iy)u(p_{2}),\label{A1}\\
-i\mathcal{M}_{A}  &  =\overline{u}(p_{1}^{\prime})(ie_{3}\gamma^{\mu}%
)u(p_{1})\left[  \langle A_{\mu}A_{\nu}\rangle\right]  \overline{u}%
(p_{2}^{\prime})(ie_{3}\gamma^{\nu})u(p_{2}), \label{A2}%
\end{align}
Here, $\langle\varphi\varphi\rangle$ and $\langle A_{\mu}A_{\nu}\rangle$ are
obviously the scalar and photon propagators given in eqs. (\ref{Prop_gauge}),
(\ref{Prop_scalar}). The scattering amplitudes, of eqs. (\ref{A1}) and
(\ref{A2}), are written for electrons of equal polarization mediated by the
scalar and gauge particles, respectively. The spinors $u(p)$ stand for the
positive-energy solution of the Dirac equation $\left(
\rlap{\hbox{$\mskip1 mu /$}}p-m\right)  u(p)=0$. The $\gamma-$ matrices
satisfy the $so(1,2)$ algebra, $\left[  \gamma^{\mu},\gamma^{\nu}\right]
=2i\epsilon^{\mu\nu\alpha}\gamma_{\alpha}$, and correspond to the Pauli
matrices: $\gamma^{\mu}=(\sigma_{z},-i\sigma_{x},i\sigma_{y}).$ Considering
all it, the following spinors%

\begin{equation}
u(p)=\frac{1}{\sqrt{N}}\left[
\begin{array}
[c]{c}%
E+m\\
-ip_{x}-p_{y}%
\end{array}
\right]  ,\text{ \ \ }\overline{u}(p)=\frac{1}{\sqrt{N}}\left[
\begin{array}
[c]{cc}%
E+m & -ip_{x}+p_{y}%
\end{array}
\right]  , \label{spinor}%
\end{equation}
are explicitly obtained. They satisfy the normalization condition
$\overline{u}(p)u(p)=1,$ for $N=2m(E+m)$. The M\"{o}ller scattering is easily
attained in the frame of the center of mass, where the momenta of the incoming
and outgoing electrons are read in the form:%
\[
p_{1}^{\mu}=(E,p,0),p_{2}^{\mu}=(E,-p,0),p_{1}^{\prime\mu}=(E,p\cos
\theta,p\sin\theta),p_{2}^{\prime\mu}=(E,-p\cos\theta,-p\sin\theta).
\]
\ The transfer 4-momentum, carried by the gauge or scalar mediators, is:
$k^{\mu}=p_{1}^{\mu}-p_{1}^{\prime\mu}=(0,p(1-\cos\theta),-p\sin\theta),$
whereas $\theta$ is the scattering angle (in the CM frame).

Starting from eqs. (\ref{Prop_scalar}), (\ref{A1}) and from the definitions
above, the scattering amplitude associated with the scalar intermediation is
readily written:
\begin{equation}
\mathcal{M}_{scalar}=y^{2}\frac{\boxplus(k)}{\boxtimes(k)(k^{2}-M_{A}^{2})}.
\label{Mscalar1}%
\end{equation}
In the case of a purely timelike background, $v^{\mu}=($v$_{0}$,
$\overrightarrow{0}),$ and considering the general expression for the transfer
momentum, $k^{\mu}=(0,\mathbf{k})$, this amplitude takes the following form:%

\begin{equation}
\mathcal{M}_{scalar}=-y^{2}\frac{[\mathbf{k}^{4}+(2M_{A}^{2}+s^{2}%
)^{2}\mathbf{k}^{2}+M_{A}^{4}]}{[\mathbf{k}^{4}+\left(  2M_{A}^{2}%
+s^{2}-\text{v}_{0}^{2}\right)  \mathbf{k}^{2}+M_{A}^{4}][\mathbf{k}^{2}%
+M_{A}^{2}]}, \label{Mscalar2}%
\end{equation}
whose Fourier transform will lead to the potential that reflects the scalar
interaction carried by the $\varphi-$field.

One the other hand, in the case of the gauge intermediation, the situation is
more complicated as a consequence of the eleven terms present in the
propagator (\ref{Prop_gauge}). However, as a consequence of the
current-conservation law ($k_{\mu}J^{\mu}=0),$ only six terms of the gauge
propagator contribute to the scattering amplitude ($\theta^{\mu\nu},S^{\mu\nu
},\Lambda^{\mu\nu},T^{\mu}T^{\nu},Q^{\mu\nu},Q^{\nu\mu}).$ The first two terms
provide, in the non-relativistic limit, the Maxwell-Chern-Simons-Proca
(MCSP)\ scattering amplitude, which leads to MCSP potential. The
non-relativistic current-current amplitudes involving these two terms,
\begin{align}
j^{\mu}(p_{1})(\theta_{\mu\nu})j^{\nu}(p_{2}) &  =1,\\
j^{\mu}(p_{1})(S_{\mu\nu})j^{\nu}(p_{2}) &  =\mathbf{k}^{2}/m_{e}%
-(2i/m_{e})\mathbf{k}\times\mathbf{p,}%
\end{align}
are evaluated in refs.\cite{MCS}, \cite{Tese},\cite{PRD2}. The corresponding
scattering amplitude is then given by:
\begin{equation}
\mathcal{M}_{\theta S}=e_{3}^{2}\left\{  \left(  \frac{(\mathbf{k}^{2}%
+M_{A}^{2})}{\boxplus(k)}\right)  -\frac{s}{\boxplus}\left[  \frac
{\mathbf{k}^{2}}{m_{e}}-\frac{2i}{m_{e}}i\mathbf{k}\times\mathbf{p}\right]
\right\}  .\label{MCSP}%
\end{equation}

The current-current amplitude associated with the other terms of the gauge
potential are also carried out, assuming the form below:
\begin{align}
j^{\mu}(p_{1})(T_{\mu}T_{\nu})j^{\nu}(p_{2})  &  =-2\frac{\mathbf{p}^{4}%
}{m_{e}}\text{v}_{0}^{2}e^{i\theta}[1-\cos\theta];\\
j^{\mu}(p_{1})\text{ }(\Lambda_{\mu\nu})j^{\nu}(p_{2})  &  =\text{v}_{0}%
^{2};\\
j^{\mu}(p_{1})\text{ }(Q_{\mu\nu}-Q_{\nu\mu})j^{\nu}(p_{2})  &  =2\frac
{\mathbf{p}^{2}}{m_{e}}\text{v}_{0}^{2}[1-\cos\theta-i\sin\theta].
\end{align}

\bigskip These terms lead to the following scattering amplitudes:
\begin{align}
\text{ }\mathcal{M}_{TT}  &  =0,\mathcal{M}_{\Lambda}=-\frac{e_{3}^{2}%
s^{2}\text{v}_{0}^{2}\mathbf{k}^{4}}{(\mathbf{k}^{2}+M_{A}^{2})\boxtimes
(k)\boxplus(k)},\label{M2}\\
\mathcal{M}_{QQ}  &  =-e_{3}^{2}s\text{v}_{0}^{2}\frac{\mathbf{k}^{2}%
}{\boxtimes(k)\boxplus(k)}\frac{2}{m_{e}}\left\{  \mathbf{k}^{2}%
-2i\mathbf{k}\times\mathbf{p}\right\}  , \label{MQQ}%
\end{align}
where $\overrightarrow{p}$ $=\frac{1}{2}(\overrightarrow{p}_{1}%
-\overrightarrow{p}_{2})$ is defined in terms of the momenta $\overrightarrow
{p}_{1},\overrightarrow{p}_{2}$ of the incoming electrons and $\mathbf{p}%
^{2}=\mathbf{k}^{2}/[2(1-\cos\theta)].$ The total current-current amplitude
mediated by the massive gauge particle corresponds to the sum of four
contributions,
\begin{equation}
\mathcal{M}_{gauge}=\mathcal{M}_{\theta S}+\text{ }\mathcal{M}_{\Lambda
}+\mathcal{M}_{TT}+\mathcal{M}_{QQ}, \label{Mtotal}%
\end{equation}
where the terms $\mathcal{M}_{\Lambda},\mathcal{M}_{TT},$ and $\mathcal{M}%
_{QQ}$ lead to\thinspace background-depending corrections to the
MCS-amplitude. Notice that the amplitude $\mathcal{M}_{TT}$ was taken as null
due to its dependence on $\mathbf{p}^{4}$(working in the nonrelativistic
approximation, $p^{2}\ll m^{2})$.

\section{The electron-electron interaction potential}

\subsection{The scalar potential}

\begin{figure}
[ptb]
\begin{center}
\includegraphics[
height=3.039in,
width=3.039in
]%
{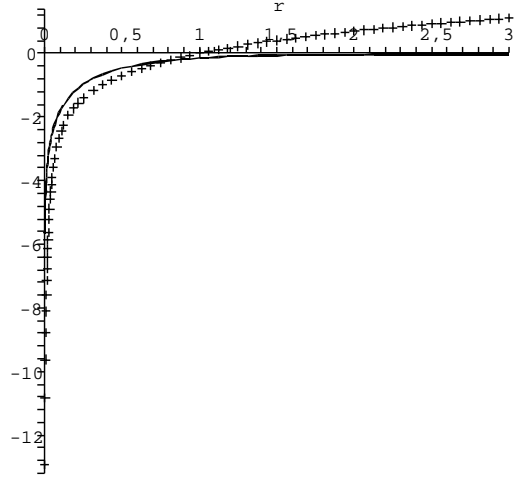}%
\caption{Simultaneous plot of the scalar potential of eq. (\ref{Vscalar})
(continous line) and the scalar potential of ref. \cite{PRD1} (cross dotted
line). Here, for both plots it was used: $s=20eV,M_{A}=2eV,$v$_{0}=2eV$.}%
\label{Scalarplot1}%
\end{center}
\end{figure}

Firstly, it is necessary to evaluate the interaction related with the scalar
intermediation. According to the Born approximation, the scalar interaction
potential is given by the Fourier transform of the scattering amplitude
(\ref{Mscalar2}), that is:%
\begin{equation}
V_{scalar}(r)=-\frac{y^{2}}{\left(  2\pi\right)  ^{2}}\int\left[
\frac{[\mathbf{k}^{4}+(2M_{A}^{2}+s^{2})^{2}\mathbf{k}^{2}+M_{A}^{4}%
]}{[\mathbf{k}^{4}+\left(  2M_{A}^{2}+s^{2}-\text{v}_{0}^{2}\right)
\mathbf{k}^{2}+M_{A}^{4}][\mathbf{k}^{2}+M_{A}^{2}]}\right]  e^{i\mathbf{k}%
\cdot\mathbf{r}}d^{2}\mathbf{k}%
\end{equation}
In this form, this integral can not be exactly solved. However, it is possible
to factorize the integrand in small factors so that an exact integration
becomes feasible. In this sense, it is important to note that:~$[\mathbf{k}%
^{4}+\left(  2M_{A}^{2}+s^{2}-\text{v}_{0}^{2}\right)  \mathbf{k}^{2}%
+M_{A}^{4}]=[\mathbf{k}^{2}+M_{+}^{2}][\mathbf{k}^{2}+M_{-}^{2}],$ where the
constants $M_{\pm}^{2}$ are given as below. After some algebraic calculations,
one obtains
\[
\frac{\lbrack\mathbf{k}^{4}+(2M_{A}^{2}+s^{2})^{2}\mathbf{k}^{2}+M_{A}^{4}%
]}{[\mathbf{k}^{4}+\left(  2M_{A}^{2}+s^{2}-\text{v}_{0}^{2}\right)
\mathbf{k}^{2}+M_{A}^{4}][\mathbf{k}^{2}+M_{A}^{2}]}=-\frac{(G+D)}%
{[\mathbf{k}^{2}+M_{+}^{2}]}+\frac{(G+E)}{[\mathbf{k}^{2}+M_{-}^{2}]}%
+\frac{(1+D-E)}{[\mathbf{k}^{2}+M_{A}^{2}]},
\]
with coefficients and mass parameters given as:%
\begin{align}
C  &  =\left[  1/\sqrt{(s^{2}-\text{v}_{0}^{2})(s^{2}-\text{v}_{0}^{2}%
+4M_{A}^{2})}\right]  ,G=\text{v}_{0}^{2}C,\text{ \ }D=G\alpha_{+},\text{
\ }E=G\alpha_{-},\label{b1}\\
\alpha_{\pm}  &  =\frac{2M_{A}^{2}}{(s^{2}-\text{v}_{0}^{2})\pm\sqrt
{(s^{2}-\text{v}_{0}^{2})(s^{2}-\text{v}_{0}^{2}+4M_{A}^{2})}},\label{t1}\\
M_{\pm}^{2}  &  =\frac{1}{2}\left[  (s^{2}-\text{v}_{0}^{2}+2M_{A}^{2}%
)\pm\sqrt{(s^{2}-\text{v}_{0}^{2})(s^{2}-\text{v}_{0}^{2}+4M_{A}^{2})}\right]
. \label{M1}%
\end{align}
Performing the Fourier transforms, the expression \
\begin{equation}
V_{scalar}(r)=\frac{y^{2}}{\left(  2\pi\right)  }\left\{  (G+D)K_{0}%
(M_{+}r)-(G+E)K_{0}(M_{-}r)-(1+D-E)K_{0}(M_{A}r)\right\}  \label{Vscalar}%
\end{equation}
is straightforwardly obtained. This result reveals a totally screened
potential as a consequence of the massive character of the scalar
intermediation. Near the origin, this potential behaves as a pure logarithm,
that is:%

\begin{equation}
\lim_{r\rightarrow0}V_{scalar}(r)=\frac{y^{2}}{\left(  2\pi\right)  }\ln r,
\end{equation}
whence one reaffirms its attractive character at the origin. Far from the
origin this potential vanishes exponentially, and is the point where it
differs from the scalar potential obtained in the Lorentz-violating
MCS\footnote{The scalar potential has the form: $V_{scalar}(r)=-\frac{y^{2}%
}{\left(  2\pi\right)  }\left\{  \left[  1+\frac{s^{2}}{w^{2}}\right]
K_{0}(sr)-\frac{s^{2}}{w^{2}}\ln r\right\}  ,$ whereas the gauge potential
derived in this paper is: $V_{gauge}(r)=\frac{e_{3}^{2}}{\left(  2\pi\right)
}\biggl\{-2(s/m)K_{0}(sr)+[s/m+s^{2}/w^{2}]K_{0}(wr)+\left(  \text{v}_{0}%
^{2}/w^{2}\right)  \ln r$ $-\frac{2}{ms}\frac{l}{r^{2}}\left[  (1+\text{v}%
_{0}^{2}/w^{2})-(s^{2}/w)rK_{1}(sr)\right]  \biggr\}.$} case \cite{PRD1},
which exhibits an asymptotic confining logarithmic behavior. It is well-known
that such kind of confining potential can not describe a physical interaction
in (1+2) dimensions. Hence, the first advantage arising from the introduction
of the Higgs sector in this theoretical framework is the transformation of the
non physical confining potential of ref. \cite{PRD1} in a Bessel\ $K_{0}$
potential (entirely suitable for describing a planar interaction). The graphic
in Fig. \ref{Scalarplot1} illustrates such a change of asymptotic behavior.%

%TCIMACRO{\FRAME{ftbpFU}{3.039in}{3.039in}{0pt}{\Qcb{Simultaneous plot of the
%scalar potential of eq. (\ref{Vscalar}) (continous line) and the scalar
%potential of ref. \cite{PRD1} (cross dotted line). Here, for both plots it was
%used: $s=20eV,M_{A}=2eV,$v$_{0}=2eV$.}}{\Qlb{Scalarplot1}}{scalarplot1.eps}%
%width 3.039in;  height 3.039in;  depth 0pt;  original-width 5.5564in;
%original-height 5.5564in;  cropleft "0";  croptop "1";  cropright "1";
%cropbottom "0";  filename 'Scalarplot1.eps';file-properties "XNPEU";}}}%
%BeginExpansion
%EndExpansion

\subsection{The gauge potential}

The starting point for the evaluation of the gauge potential is the attainment
of the $V_{\theta S}-$potential from the Fourier transform of the
$\mathcal{M}_{\theta S}$-amplitude. Such Fourier transform can not be directly
solved from eq. (\ref{MCSP}), which then is properly factorized in the form%

\begin{equation}
\mathcal{M}_{\theta S}=e_{3}^{2}\left\{  \frac{A_{+}}{[\mathbf{k}^{2}%
+m_{+}^{2}]}+\frac{A_{-}}{[\mathbf{k}^{2}+m_{-}^{2}]}+\left[  \frac
{B}{[\mathbf{k}^{2}+m_{+}^{2}]}-\frac{B}{[\mathbf{k}^{2}+m_{-}^{2}]}\right]
\left(  \frac{\mathbf{k}^{2}}{m_{e}}-\frac{2i}{m_{e}}i\mathbf{k}%
\times\mathbf{p}\right)  \right\}  , \label{MCSP2}%
\end{equation}
where:
\begin{equation}
A_{\pm}=\frac{1}{2}(1\pm s/\sqrt{s^{2}+4M_{A}^{2}}),\text{ }B=1/(\sqrt
{s^{2}+4M_{A}^{2}}),\text{ }m_{\pm}^{2}=\frac{1}{2}\left[  (s^{2}+2M_{A}%
^{2})\pm s\sqrt{s^{2}+4M_{A}^{2})}\right]  .
\end{equation}
Carrying out the Fourier transform of eq. (\ref{MCSP2}), the following
potential turns out:
\begin{equation}
V_{\theta S}(r)=\frac{e_{3}^{2}}{\left(  2\pi\right)  }\biggl\{[A_{+}-\frac
{B}{m_{e}}m_{+}^{2}]K_{0}(m_{+}r)+[A_{-}+\frac{B}{m_{e}}m_{-}^{2}]K_{0}%
(m_{-}r)-\frac{2Bl}{m_{e}r}[m_{+}K_{1}(m_{+}r)-m_{-}K_{1}(m_{-}r)]\biggr\},
\end{equation}
where $l=\overrightarrow{r}\times\overrightarrow{p}$ is the angular momentum
(a scalar in a two-dimensional space). It is interesting to point out that
$V_{\theta S}$ is exactly the electron-electron MCS-Proca potential, obtained
firstly in ref. \cite{Tese}. Near the origin, one has: $K_{0}(mr)\rightarrow
-\ln r,$ $K_{1}(mr)\rightarrow1/(mr)+mr(\ln r)/2,$ implying the following
result:%
\begin{equation}
\lim_{r\rightarrow0}V_{\theta S}(r)\rightarrow-\frac{e_{3}^{2}}{\left(
2\pi\right)  }[1-\frac{s}{2m_{e}}(1+l)]\ln r. \label{MCSP_limit}%
\end{equation}
Once one works in the limit of small topological mass, $s<<m_{e},$ this
potential exhibits a repulsive behavior near the origin.

The interaction potential associated with the amplitudes $\mathcal{M}%
_{\Lambda}\ $may be obtained from the Fourier transform of the scattering
amplitude given in eq. (\ref{M2}); however, such amplitude must be previously
factorized as%

\[
\mathcal{M}_{\Lambda}=e_{3}^{2}s^{2}\biggl\{\left[  -C(L_{+}-L_{-}%
)+h(l_{+}-l_{-})\right]  \frac{1}{[\mathbf{k}^{2}+M_{A}^{2}]}-(CN_{+}%
-hn_{+})\frac{1}{[\mathbf{k}^{2}+M_{+}^{2}]}+(CN_{-}-hn_{-})\frac
{1}{[\mathbf{k}^{2}+M_{-}^{2}]}\biggr\},
\]
with the coefficients given as:
\begin{equation}
L_{\pm}=\frac{-M_{A}^{2}}{M_{\pm}^{2}-M_{A}^{2}},\text{ }l_{\pm}=\frac
{-M_{A}^{2}}{m_{\pm}^{2}-M_{A}^{2}},N_{\pm}=\frac{M_{+}^{2}}{M_{\pm}^{2}%
-M_{A}^{2}},\text{ }n_{\pm}=\frac{m_{+}^{2}}{m_{\pm}^{2}-M_{A}^{2}},h=B/s.
\end{equation}
The Fourier transforms are then performed, leading to a combination of $K_{0}$
functions, namely:%

\begin{equation}
V_{\Lambda}(r)=\frac{e_{3}^{2}s^{2}}{\left(  2\pi\right)  }\biggl\{\left[
-C(L_{+}-L_{-})+h(l_{+}-l_{-})\right]  K_{0}(M_{A}r)-(CN_{+}-hn_{+}%
)K_{0}(M_{+}r)+(CN_{-}-hn_{-})K_{0}(M_{-}r)\biggr\},
\end{equation}
Near the origin, this potential vanishes identically, $\lim_{r\rightarrow
0}V_{\Lambda}(r)\rightarrow0;$ far from the origin, it decays exponentially.

Applying the same procedure to $\mathcal{M}_{QQ}$, after some algebra, it ends
up in:%

\[
\mathcal{M}_{QQ}=-\frac{e_{3}^{2}s}{m_{e}}\biggl\{\left[  \frac{-C}%
{[\mathbf{k}^{2}+M_{+}^{2}]}+\frac{C}{[\mathbf{k}^{2}+M_{-}^{2}]}+\frac
{h}{[\mathbf{k}^{2}+m_{+}^{2}]}-\frac{h}{[\mathbf{k}^{2}+m_{-}^{2}]}\right]
(\mathbf{k}^{2}-2i\mathbf{k}\times\mathbf{p})\biggr\},
\]
so that the resulting potential is:%

\begin{align}
V_{QQ}(r)  &  =\frac{e_{3}^{2}}{\left(  2\pi\right)  }\frac{s}{m_{e}%
}\biggl\{\frac{2l}{r}[-C(M_{+}K_{1}(M_{+}r)-M_{-}K_{1}(M_{-}r))+h(m_{+}%
K_{1}(m_{+}r)-m_{-}K_{1}(m_{-}r))]\nonumber\\
&  -C[M_{+}^{2}K_{0}(M_{+}r)-M_{-}^{2}K_{0}(M_{-}r)]+h[m_{+}^{2}K_{0}%
(m_{+}r)-m_{-}^{2}K_{0}(m_{-}r)]\biggr\}.
\end{align}
This latter potential exhibits the same behavior of $V_{\Lambda}$ near and far
from the origin, that is: $\ \lim_{r\rightarrow0,\infty}V_{QQ}(r)\rightarrow
0.$

The total gauge interaction potential, $V_{gauge}(r)=V_{\theta S}+V_{\Lambda
}+V_{QQ},$ after some simplifications, assumes the explicit form:%

\begin{align}
V_{gauge}(r)  &  =\frac{e_{3}^{2}}{\left(  2\pi\right)  }\biggl\{A_{+}%
K_{0}(m_{+}r)+A_{-}K_{0}(m_{-}r)-[s^{2}(CN_{+}-hn_{+})+C\frac{s}{m_{e}}%
M_{+}^{2}]K_{0}(M_{+}r)\nonumber\\
&  +[s^{2}(CN_{-}-hn_{-})+CM_{-}^{2}s/m_{e}]K_{0}(M_{-}r)+s^{2}\left[
-C(L_{+}-L_{-})+h(l_{+}-l_{-})\right]  K_{0}(M_{A}r)\nonumber\\
&  +\frac{2l}{r}\frac{s}{m_{e}}C[M_{+}K_{1}(M_{+}r)+M_{-}K_{1}(M_{-}%
r)]\biggr\}. \label{Vgauge1}%
\end{align}

This full expression corresponds to the MCS-Proca potential ($V_{\theta S})$
corrected by the Lorentz-violating terms arising from $V_{\Lambda},V_{QQ}.$ In
the limit of a vanishing background (v$_{0}\rightarrow0),$ one has
$V_{\Lambda},V_{QQ}$ $\rightarrow0$, remaining only the $V_{\theta S}$
potential, which shows the consistency of the obtained results. Obviously,
this is an expected outcome, since both $V_{\Lambda},V_{QQ}$ are potential
contributions induced merely by the presence of the background.

\begin{figure}
[ptb]
\begin{center}
\includegraphics[
height=3.039in,
width=3.039in
]%
{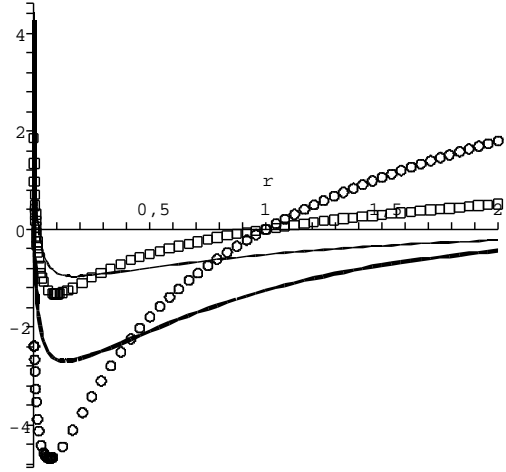}%
\caption{Simultaneous plot of the gauge potential of eq. (\ref{Vgauge1})
(continuos line) and the gauge potential of ref. \cite{PRD1} (dotted line) for
two sets of values: $l=1,s=20,m_{e}=10^{5},M_{A}=3,$v$_{0}=13eV$ (continuos
thin line and box dotted curve); $l=1,s=20,m_{e}=10^{5},M_{A}=3,$v$_{0}=17eV$
(continuos thicker line and circle dotted curve).}%
\label{Gauge_gauge}%
\end{center}
\end{figure}

Far from the origin, this potential vanishes exponentially (according to the
asymptotic behavior of the Bessel functions), a consequence of the massive
character of the physical mediators. In this point, this outcome differs from
the asymptotic logarithmically divergent gauge potential attained in ref.
\cite{PRD1} (which is written in footnote 1). The graph of Fig.
[\ref{Gauge_gauge}] shows a simultaneous plot of the gauge potential of eq.
(\ref{Vgauge1}) (continuos line) and the one of ref. \cite{PRD1} (dotted
line), which has a deeper minimum, compared with the former (for each set of parameters).%

%TCIMACRO{\FRAME{ftbpFU}{3.039in}{3.039in}{0pt}{\Qcb{Simultaneous plot of the
%gauge potential of eq. (\ref{Vgauge1}) (continuos line) and the gauge
%potential of ref. \cite{PRD1} (dotted line) for two sets of values:
%$l=1,s=20,m_{e}=10^{5},M_{A}=3,$v$_{0}=13eV$ (continuos thin line and box
%dotted curve); $l=1,s=20,m_{e}=10^{5},M_{A}=3,$v$_{0}=17eV$ (continuos thicker
%line and circle dotted curve).}}{\Qlb{Gauge_gauge}}{gauge_gaugeplot1.eps}%
%width 3.039in;  height 3.039in;  depth 0pt;  original-width 5.5564in;
%original-height 5.5564in;  cropleft "0";  croptop "1";  cropright "1";
%cropbottom "0";  filename 'Gauge_gaugeplot1.eps';file-properties "XNPEU";}}}%
%BeginExpansion
%EndExpansion

\begin{figure}
[ptb]
\begin{center}
\includegraphics[
height=3.0381in,
width=3.0381in
]%
{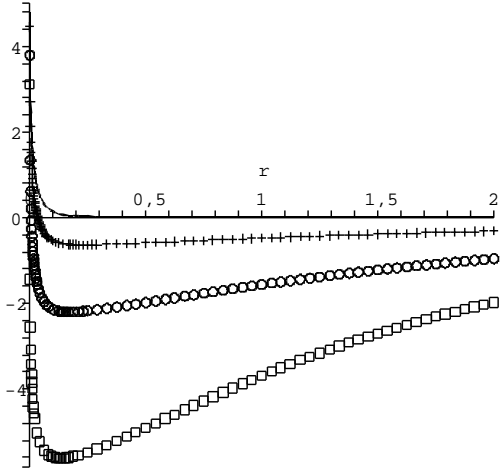}%
\caption{Simultaneous plot of the gauge potential for four sets of parameter
values: $m_{e}=10^{5},l=1,$ $s=20eV,M_{A}=2eV,$v$_{0}=0$ (continuos line)$;$
v$_{0}=10$ $eV$ (cross dotted line); v$_{0}=15$ $eV$ (circle dotted line);
v$_{0}=18$ $eV$ (box dotted line).}%
\label{Gauge1plot1}%
\end{center}
\end{figure}

Near the origin, the Lorentz-violating contributions of eq. (\ref{Vgauge1})
tend to zero, so that in this limit the gauge potential is entirely ruled by
the $V_{\theta S}$ contribution, namely:%

\begin{equation}
V_{gauge}(r)\simeq\biggl\{-\frac{e_{3}^{2}}{\left(  2\pi\right)  }[1-\frac
{s}{2m_{e}}(1+l)]\ln r\biggr\}. \label{Vgauge2}%
\end{equation}
It is interesting to note that this is the same behavior of the gauge
potentials achieved in refs. \cite{PRD1}, \cite{PRD2}. As already claimed,
this potential will always exhibit a repulsive behavior near the origin. This
general behavior is illustrated in Fig.[\ref{Gauge1plot1}] for four sets of parameters.%

%TCIMACRO{\FRAME{ftbpFU}{3.0381in}{3.0381in}{0pt}{\Qcb{Simultaneous plot of the
%gauge potential for four sets of parameter values: $m_{e}=10^{5},l=1,$
%$s=20eV,M_{A}=2eV,$v$_{0}=0$ (continuos line)$;$ v$_{0}=10$ $eV$ (cross dotted
%line); v$_{0}=15$ $eV$ (circle dotted line); v$_{0}=18$ $eV$ (box dotted
%line).}}{\Qlb{Gauge1plot1}}{gauge1plot1.eps}%
%width 3.0381in;  height 3.0381in;  depth 0pt;  original-width 5.5564in;
%original-height 5.5564in;  cropleft "0";  croptop "1";  cropright "1";
%cropbottom "0";  filename 'Gauge1plot1.eps';file-properties "XNPEU";}}}%
%BeginExpansion
%EndExpansion

Concerning such a picture, it presents a comparison of the electron-electron
MCS-Proca potential (corresponding to the case for which v$_{0}=0)$ with the
gauge one for three different values of v$_{0}$. It shows that the gauge
potential appreciably deviates from the MCS-Proca behavior as the larger is
the magnitude of the background (v$_{0})$, that is, the larger the background
modulus the deeper the attractive region of the potential. The attractiveness
of the gauge potential is ascribed to the presence of the well-shaped region
(constituted by a part of decreasing behavior followed by a part of increasing
behavior), exhibited in the graphics of Fig. 2. In a dynamic perspective, such
a well-shaped curve may be described in terms of a region in which the
gradient potential is negative followed by a positive gradient region in the
sequel. Another\ kind of potential curve which implies an attractive behavior
is one whose potential gradient is always negative (but with a decreasing
modulus with increasing distance, $|\nabla V|\rightarrow0$ for $r\rightarrow
\infty).$ This is the behavior exhibited by the scalar potential in
Fig.[\ref{Scalarplot1}].

The discussion on the attractiveness of the gauge potential must be conducted
with caution and can not be based only on the expression contained in eq.
(\ref{Vgauge1}). It happens that, specifically in (1+2) dimensions, a
tree-level result may be altered by the 1-loop contributions associated with
2-photon diagrams. This fact was put in evidence by the controversy involving
the attractive/repulsive character of the MCS potential \cite{MCS}, which has
shown that this potential turns out truly repulsive (instead of attractive)
whenever the 2-photon diagrams that assure its gauge invariance are taken into
account. In short, such a discussion has shown that the correct behavior of a
(1+2)- potential can only be achieved if the 2-photons diagrams are
considered. Nevertheless, there is a way to circumvent the awkward calculation
of such diagrams, which consists in requiring the gauge invariance of the
Pauli equation,
\begin{equation}
\left[  \frac{(\mathbf{p}-e_{3}\mathbf{A})^{2}}{m_{e}}+e\phi(r)-\frac
{\mathbf{\sigma\cdot B}}{m_{e}}\right]  \Psi(r,\phi)=E\Psi(r,\phi),
\label{Pauli}%
\end{equation}
and keeping it in the non-relativistic limit (governed by the Schr\"{o}dinger
equation). The gauge invariance of Pauli equation is assured by the presence
of the $A^{2}-$term, which obviously does not appear\thinspace in the context
of a nonperturbative low-energy evaluation (once it is associated with
2-photon exchange processes), but may come to be as relevant as the tree-level
ones (see Hagen and Dobroliubov \cite{MCS}) in (1+2) dimensions. Therefore,
both this term and the centrifugal barrier must be kept as correction terms of
the effective potential for the Schr\"{o}dinger equation derived from eq.
(\ref{Pauli}). It is this effective potential that represents the true
electron-electron interaction in the non-relativistivistic limit. In order to
obtain such a potential explicitly, one writes the Laplacian operator,
$\left[  \partial^{2}/\partial r^{2}+r^{-1}\partial/\partial r+r^{-2}%
\partial^{2}/\partial\phi^{2}\right]  ,$ corresponding to the $p^{2}-$term,
whose action on the total wavefunction, $\Psi(r,\phi)=R_{nl}(r)e^{i\phi l},$
generates\ the repulsive centrifugal barrier term, $l^{2}/mr^{2}.$ Such a term
is then added to $e_{3}^{2}A\cdot A/m_{e}$, already presented in eq.
(\ref{Pauli}), thus yielding the effective potential: $V_{eff}(r)=V_{gauge}%
(r)+l^{2}/\left(  mr^{2}\right)  +e_{3}^{2}(A\cdot A)/m_{e}$.

\begin{figure}
[ptb]
\begin{center}
\includegraphics[
height=3.0381in,
width=3.0381in
]%
{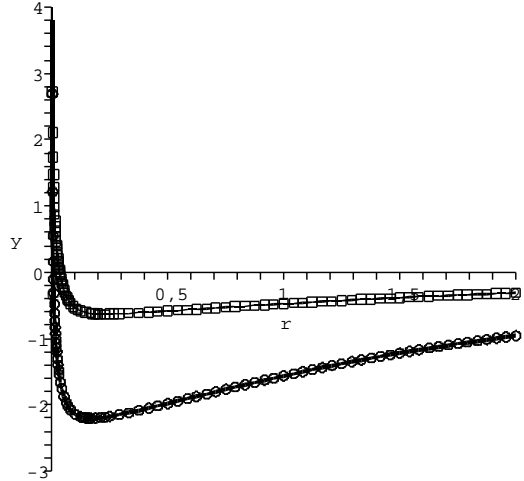}%
\caption{Simultaneous plot of the gauge potential (\ref{Vgauge1}) and the
effective potential (\ref{Veff}) for the following sets of parameter values:
$s=20eV,l=1,m_{e}=10^{5}eV,M_{A}=2eV,$v$_{0}=10eV$ (overlapped continuos line
and box dotted line); $s=20eV,l=1,m_{e}=10^{5},M_{A}=2eV,$v$_{0}=15eV$
(overlapped continuos line and circle dotted line).}%
\label{Veffplot1}%
\end{center}
\end{figure}

The vector potential, $\mathbf{A}$, stemming from the planar model described
by Lagrangian (\ref{Lagrange3}), has been already evaluated in ref.
\cite{Manojr5} for the timelike case:%
\begin{equation}
\mathbf{A}(r)=-\frac{e_{3}s}{(2\pi)}C\left[  M_{+}K_{1}\left(  M_{+}r\right)
-M_{-}K_{1}\left(  M_{-}r\right)  \right]  \overset{\wedge}{r^{\ast}},
\end{equation}
whence the effective potential takes the form:%
\begin{equation}
V_{eff}(r)=V_{gauge}(r)+\frac{l^{2}}{m_{e}r^{2}}+\left(  \frac{e_{3}}{2\pi
}\right)  ^{2}\frac{e_{3}^{2}s^{2}C^{2}}{m_{e}}[M_{+}K_{1}\left(
M_{+}r\right)  -M_{-}K_{1}\left(  M_{-}r\right)  ]^{2}. \label{Veff}%
\end{equation}
This is the gauge invariant effective potential that comprises the two-photon
contribution ($\mathbf{A\cdot A}-$term) and the centrifugal barrier term,
leading to the correct low-energy electron-electron interaction. Based upon
such full expression, one proceeds to verify whether the electron-electron
interaction may come to be attractive in some region by means of the graphical
analysis of Fig. \ref{Veffplot1}.%
%TCIMACRO{\FRAME{ftbpFU}{3.0381in}{3.0381in}{0pt}{\Qcb{Simultaneous plot of the
%gauge potential (\ref{Vgauge1}) and the effective potential (\ref{Veff}) for
%the following sets of parameter values: $s=20eV,l=1,m_{e}=10^{5}eV,M_{A}%
%=2eV,$v$_{0}=10eV$ (overlapped continuos line and box dotted line);
%$s=20eV,l=1,m_{e}=10^{5},M_{A}=2eV,$v$_{0}=15eV$ (overlapped continuos line
%and circle dotted line).}}{\Qlb{Veffplot1}}{veffplot1.eps}%
%width 3.0381in;  height 3.0381in;  depth 0pt;  original-width 5.5564in;
%original-height 5.5564in;  cropleft "0";  croptop "1";  cropright "1";
%cropbottom "0";  filename 'Veffplot1.eps';file-properties "XNPEU";}}}%
%BeginExpansion
%EndExpansion

The graph in Fig.[\ref{Veffplot1}] shows that the effective and the gauge
potential differ from each other, for the adopted parameter values, by an
absolutely negligible amount, because in\ this case, the graphics come out
perfectly superimposed (revealing that they correspond numerically to the same
value). Thus, the effective potential also exhibits an attractive behavior,
showing that the gauge interaction is really endowed with attractiveness.

In the framework of this work, the total electron-electron interaction
encompasses the gauge and the scalar contributions:\textbf{ }$V_{total}%
(r)=V_{scalar}+V_{gauge}.$\textbf{ }This total potential is attractive
whenever\textbf{ }it presents a well-shaped region. The shape of the total
potential near the origin depends on the value of the constants $y^{2}%
,e_{3}^{2}.$ If $y^{2}>e_{3}^{2}$ the total potential will present a similar
behavior to that of the scalar potential of Fig. [1], while in the case
$y^{2}<e_{3}^{2}$ it will be approximately as the gauge potential of Figs.[2,
3]. As both these forms of potential are endowed with well regions, we
conclude that the total potential may always be attractive. This constitutes a
relevant result provided it ensures the possibility of obtaining $e^{-}e^{-}$
bound states in the framework of this particular model.\ 

\section{Concluding Remarks}

In this work, it was considered the M\"{o}ller scattering in the context of
the planar Lorentz-violating Maxwell-Chern-Simons-Proca electrodynamics,
obtained from the dimensional reduction of the Abelian-Higgs
Maxell-Carroll-Field-Jackiw model \cite{Belich1}. For the case of a purely
timelike background, the interaction potential was calculated as the Fourier
transform of the M\"{o}ller amplitude (Born approximation), carried out in the
non-relativistic regime. The attained total potential presents two distinct
contributions: the attractive scalar potential (stemming form the Yukawa
exchange) and the gauge one (mediated by the MCS-Proca gauge field). The
scalar potential, as expected, is always attractive no matter if it is near or
far from the origin, and represents a totally shielded interaction. Such an
interaction may be identified with phonon exchange processes, which represent
physical excitations in several systems of interest. As for the gauge
interaction, it is composed of the repulsive MCS-Proca potential $\left(
V_{\theta S}\right)  $ corrected by background-depending contributions, which
impose relevant physical modifications. Indeed, for larger values of v$_{0},$
the gauge potential exhibits a pronounced attractive region. Both the scalar
and gauge potentials are entirely screened interactions, a consequence of the
massive mediators generated by the Higgs mechanism. This feature, in
principle, may turn these potentials suitable for describing real planar
systems of Condensed Matter physics. This is the main difference between the
interaction potentials of this work and the potentials derived in ref.
\cite{PRD1}, which present a logarithmic asymptotic behavior and are
unsuitable for representing a physical interaction in a low-energy planar
system. Hence, in an attempt of accounting for a real interaction in a
Lorentz-violating planar system, one should adopt the potentials here derived
instead of the ones of ref. \cite{PRD1}.

In this work, one has argued that the total interaction potential exhibits an
attractive region, able to bring about the formation of $\ e^{-}e^{-}$ pairs.
As a forthcoming feasible application, one can explicitly evaluate the
$e^{-}e^{-}$ binding energies by means of the numerical solution of the
Schr\"{o}dinger equation written for the potentials here derived. This may be
done by ascribing reasonable values for the free parameters of this model, in
a similar procedure to the one of refs. \cite{Tese}, \cite{Tese2}. It is
expected that a fine tune of the parameters would yield binding energies in
the scale of $10^{-3}eV$, a typical energy for electron-electron pairing in
planar systems.

It is also important to emphasize the difference between the total potential
obtained in this work and the ones of ref. \cite{Tese}, which consist of an
always repulsive MCS-Proca contribution added to an attractive scalar
potential. In this latter case, the attractive scalar potential arises from
the intermediation played by the Higgs field, and the total potential exhibits
attractiveness only if this scalar contribution overcomes the MCS-Proca one.
Therefore, the possibility of attaining an attractive interaction depends
entirely on the presence of the Higgs mode. This is not the case of the
present work, in which the gauge potential itself may be attractive (even for
small values of \ v$_{0}$ compared to the electron rest mass) - see Figs.
[\ref{Gauge_gauge}], [\ref{Gauge1plot1}] - an effect of the background on the
system. Furthermore,\textbf{ }the scalar intermediation field is the one
stemming from the dimensional reduction ($\varphi=A^{(3)})$ instead of the
Higgs field, which now accounts for the screened character of the
interactions. \textbf{ }These are particularities that distinguish the present
model from the ones of refs. \cite{Tese}, \cite{Tese2},\cite{PRD1}.

The general connection between noncommutativity and Lorentz violation turns a
sensible matter the comparison of the Ghosh potentials \cite{Ghosh} and the
results of ref. \cite{PRD1}, in which it was evaluated the Lorentz-violating
version of the MCS potential. Yet, the expressions of these potentials result
to be different. Indeed, while the Lorentz-violating potentials increase
logarithmically with distance, the noncommutative potentials exhibit a
$1/r^{2}$ asymptotic behavior. So, it is clear that these potentials differ
from each other substantially, which justifies this investigation in the
presence of both noncommutativity and Lorentz violation.

It is instructive to clarify the reason to have adopted a purely timelike
background. This was done for a simplicity issue, since in this case the
interaction potential may be exactly solved (without approximations). The
physical interpretation of this background, however, is not a straightforward
matter, since v$_{0}$ may not be easily associated with any parameter of the
system. Some felling can be got observing the effect of the background on the
behavior of the system. As example, it was reported in ref. \cite{Manojr2}
that a purely timelike background modifies drastically the asymptotic behavior
of the electrical field of the MCS electrodynamics. In fact, while the pure
MCS solution presents an exponentially decaying solution, the
Lorentz-violating MCS electric field exhibits an increasing logarithmic
behavior. In this case, the background may be seen as a constant field that
annihilates the screening of the electric sector of the theory, changing its
asymptotic behavior. This property justifies the asymptotic logarithmic
behavior of the potentials of refs. \cite{PRD1}, \cite{PRD2}. A possible
continuation of this work consists in evaluating the electron-electron
potential in the case of a purely spacelike background, $v^{\mu}%
=(0,\mathbf{v}),$ standing for a privileged direction in space able to bring
about anisotropy for the solutions. The presence of anisotropy in (1+2)
dimensions is a factor that can be properly described in the framework of a
Lorentz-violating background.\textbf{ } In general, this is a more complicated
case in which the potentials may be only obtained within the approximation for
\textbf{v}$^{2}/s^{2}<<1,$ as done in ref. \cite{PRD2}. In principle, this is
a case where the background may be more clearly interpreted as an active
feature of an anisotropic condensed matter system, whose solutions exhibit an
explicit dependence on the direction stated by \textbf{v}.

\begin{acknowledgments}
\ The authors are grateful to FAPEMA\ (Funda\c{c}\~{a}o de Amparo \`{a}
Pesquisa do Maranh\~{a}o) and CNPq (Conselho Nacional de Desenvolvimento
Cient\'{\i}fico e Tecnol\'{o}gico) for financial support.
\end{acknowledgments}

\end{document}